\documentclass[a4paper,11pt,twoside]{scrartcl}
\usepackage{ILD}

\usepackage{xspace}

\title{Prospects for light Higgs measurements\\ at the 250 GeV ILC}

\ildproc{PHYS}{2023}{003}

\date{July 28, 2023}

\addauthor{Aleksander Filip \.Zarnecki}{\institute{1}}

\addinstitute{1}{Faculty of Physics, University of Warsaw, Poland\\
                          filip.zarnecki@fuw.edu.pl}

\abstract{
A light Higgs boson, with mass of the order of 100\,GeV, is still not
excluded by the existing experimental data, provided its coupling to
gauge bosons is strongly suppressed compared to a SM-like Higgs boson
at the same mass. Also other couplings of such a scalar could be very
different from the SM predictions leading to non-standard decay
paterns. Considered in the presented study is the feasibility of
direct observation of the 96\,GeV Higgs boson of N2HDM model with
dominant decays to tau lepton pairs.

\vspace*{1cm}

\begin{center}
  Talk presented at\\
  the International Workshop on Future Linear
  Colliders (LCWS 2023), 15-19 May 2023.\\
  C23-05-15.3.\\[1cm]
  This work was carried out in the framework of the ILD concept group
\end{center}
}

\newcommand{\whizard}{\textsc{Whizard}\xspace}
\newcommand{\pythia}{\textsc{Pythia}\xspace}
\newcommand{\delphes}{\textsc{Delphes}\xspace}

\graphicspath{{./logos/} {./plots/} }

\begin{document}

\titlepage


\section{Motivation}

While the existence of the Standard Model like Higgs boson with mass
of 125\,GeV has been firmly established, experimental results obtained
so far still  leave room for additional light scalar states \cite{Robens:2022zgk}.
Some deviations observed in the LHC Run 1 and Run 2 have been
interpreted as a possible existence of the new scalar with the mass of around 96\,GeV
\cite{Biekotter:2019kde,Heinemeyer:2021msz,Biekotter:2022jyr}.
Considered in the presented study are prospects for observing the new
scalar produced in the scalar-strahlung process, $ e^+e^- \to Z \;
\phi$, at the International Linear Collider (ILC) running at 250\,GeV.
Previous studies focused on establishing the decay independent
searches \cite{Wang:2020lkq} or on the $\phi \to b\bar{b}$ decay
channel \cite{Drechsel:2018mgd}, expected to be dominant in most scenarios.
For the presented study we selected one of the best-fit scenarios of
the N2HDM model proposed in \cite{Biekotter:2022jyr}, with scalar mass
of 95.7\,GeV, strongly suppressed $b\bar{b}$ branching ratio and
dominant scalar decay to tau lepton pair.
This seems to be an interesting and challenging benchmark scenario for the light
scalar searches at future Higgs factories.

\section{Analysis setup}

Event samples used in the presented study were generated using
\whizard \cite{Moretti:2001zz,Kilian:2007gr} version 3.1.1.
For the signal sample, UFO implementation of the Singlet-extended Two
Higgs doublet model (S2HDM)\footnote{S2HDM model is
equivalent to N2HDM, when the additional dark-matter candidate is
heavy. Its mass was set to 10\,TeV.} was used \cite{Biekotter:2021ovi},
modified for type IV couplings of the 2HDM.
Only the tau decay channel events were generated for the signal sample.
For the considered N2HDM scenario, production cross section at 250 GeV
is equal to about 11\% of the SM cross section for the assumed mass and
the expected scalar branching ratio to tau pairs is 41.2\% \cite{Biekotter:2022jyr}.
All background channels were simulated with the \whizard built-in SM\_CKM
model.
While the dominant background contribution is expected to come from the SM process
with the same final state, $e^+e^- \to q \bar{q} \tau \tau$, other
four-fermion final states were also considered as possible background sources. 

For the initial study we consider only one ILC beam polarisation
setting, with $-80\%$ electron and $+30\%$ positron polarisation, and
integrated luminosity of 900\,fb$^{-1}$ \cite{Bambade:2019fyw}.
The ILC beam energy profile was taken into account base on
\textsc{Circe2} parametrization and hadronisation was simulated with
the~\pythia~6~\cite{Sjostrand:2006za}. 
The fast detector simulation framework
\delphes~\cite{deFavereau:2013fsa} was used to simulate
detector response, with built-in cards for parametrisation of the ILC
detector, \texttt{delphes\_card\_ILCgen.tcl} \cite{bib:ILCgen}.

\section{Tau reconstruction}

\begin{figure}
\includegraphics[width=0.62\textwidth]{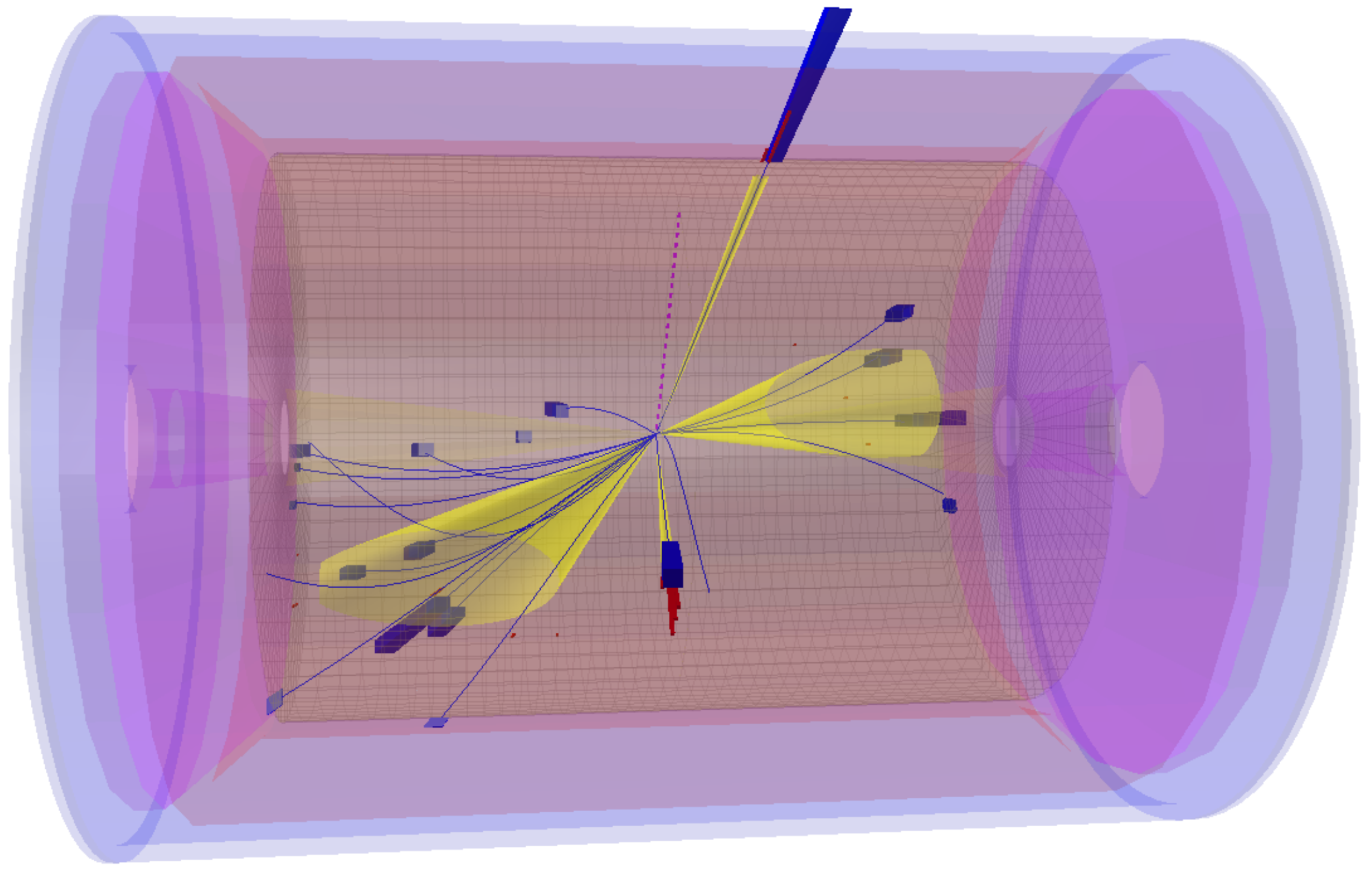}
\includegraphics[width=0.38\textwidth]{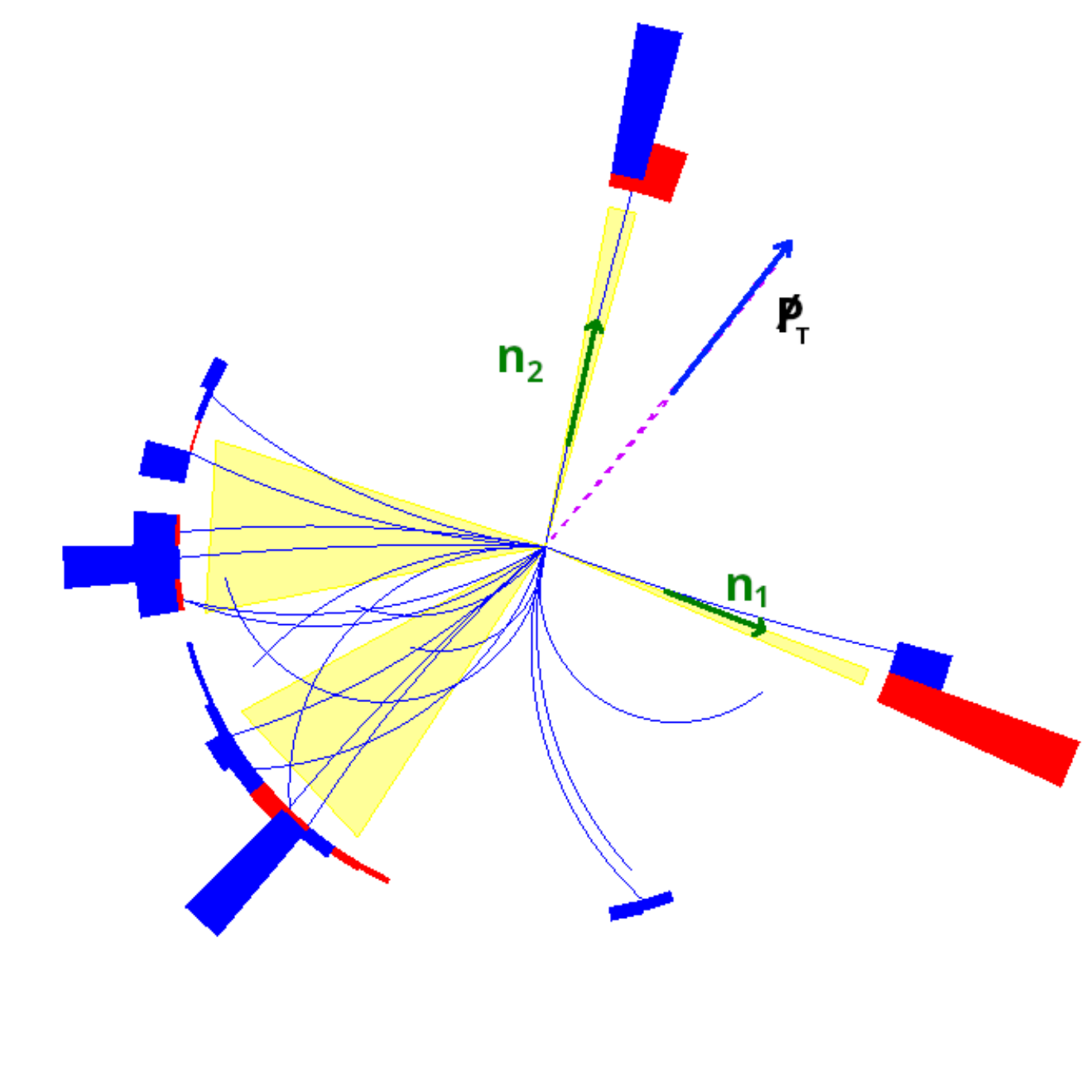}
\caption{Left: example signal event, with hadronic decays of the two tau
  leptons produced in the light scalar decay. Right: same event in the
  transverse plane, with missing transverse momentum
  p$\!\!\!\textrm{/}_\textsc{t}$ and two unit vectors along tau jet
  directions, n$_1$ and n$_2$, indicated. }
\label{event}
\end{figure}

Depending on the decays of the two tau leptons,
three decay channels can be considered for the signal events:
hadronic (with both taus decaying hadronicly), semi-leptonic (with one
leptonic tau decay) and leptonic (with leptonic decays of both taus).
Example of signal event with hadronic final state is shown in
Fig.~\ref{event}.

\begin{figure}
\includegraphics[width=0.5\textwidth]{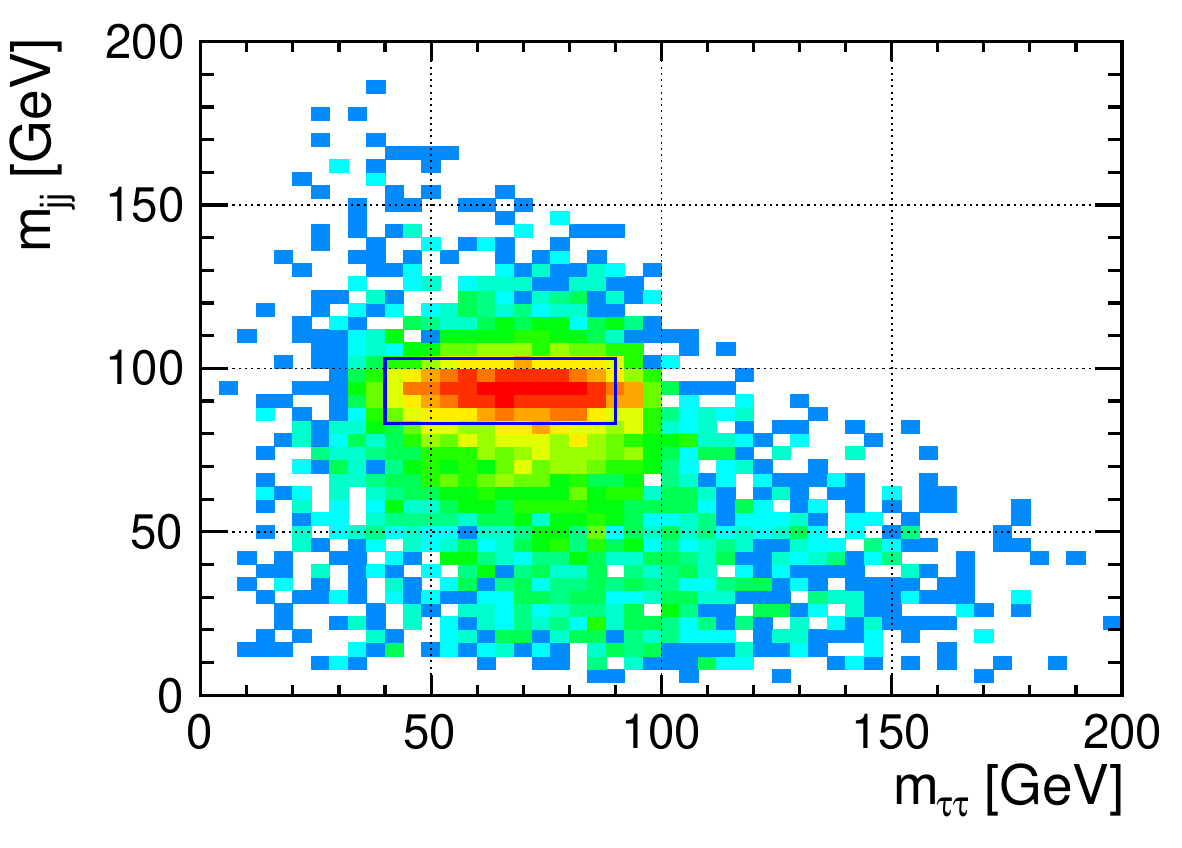}%
\includegraphics[width=0.5\textwidth]{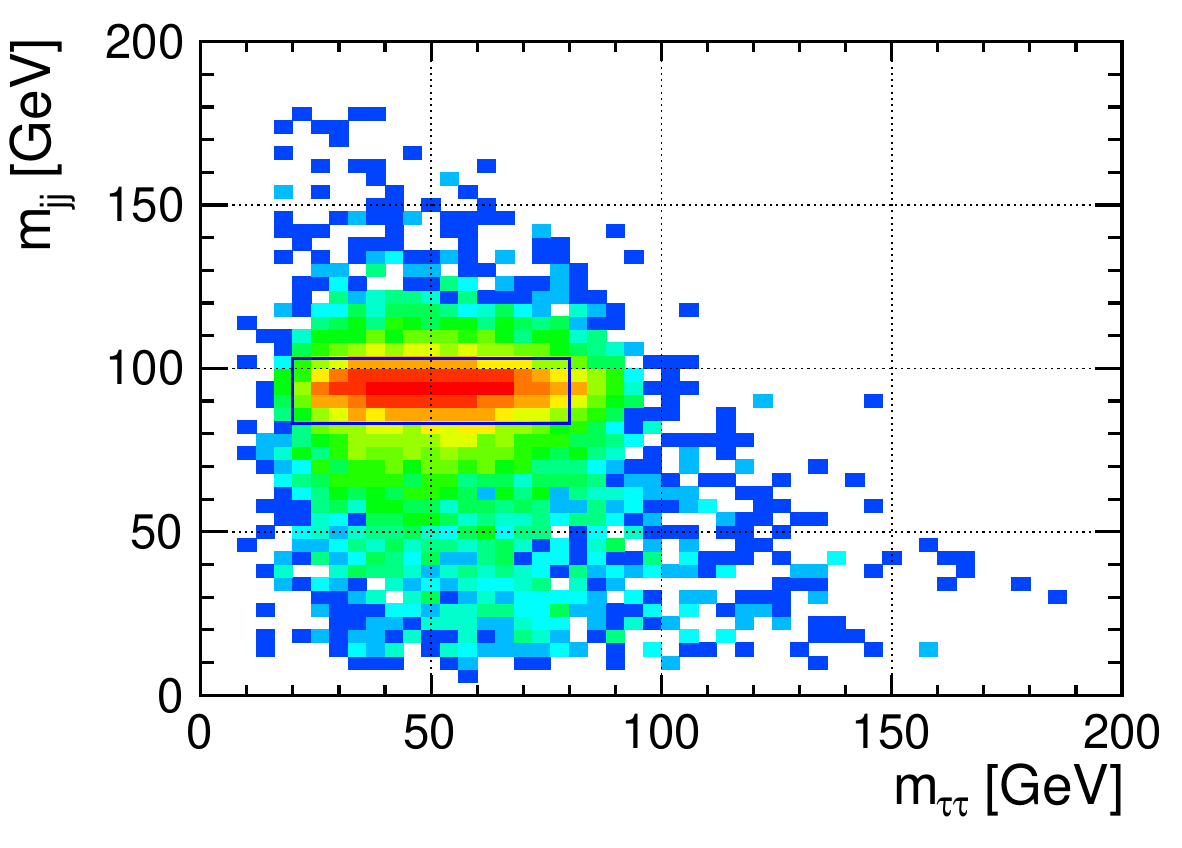}
\caption{Invariant mass of the two hadronic jets attributed to the Z
  boson decay, m$_{jj}$, as a function of the uncorrected invariant
  mass of the two tau candidates, m$_{\tau\tau}$. Compared are expected
  distributions for hadronic (left) and semi-leptonic (right) scalar
  decays. See text for details.
  } 
\label{rawmass_2d}
\end{figure}

Depending on the scalar decay channel zero, one or two isolated
leptons are expected in the final state.
Each isolated lepton (electron or muon) is considered a tau candidate.
The remaining final state is clustered into four, three or two jets,
depending on the number of leptons.
Jets are selected as coming from tau decays based on the tau tagging result.
Untagged jets are assumed to come from Z boson decay.

One of the challenges in the presented study is the reconstruction of
the invariant  mass of the two tau candidates. Distribution of the
measured invariant mass of the two jets from Z boson decay and the two tau
candidates is shown in Fig.~\ref{rawmass_2d}, for hadronic and
semi-leptonic signal events.
While the Z boson mass is quite well reconstructed for most of the
signal events, invariant mass of the tau candidate pair is
significantly underestimated due to the escaping neutrinos.

To correct for the neutrino energy, we use the so called collinear
approximation \cite{Kawada:2015wea}. For high energy tau leptons,
decay products are highly boosted in the initial lepton direction. One
can therefore assume that the initial tau lepton, escaping neutrino
and the observed tau candidate are collinear.
Neutrino energies can be found from transverse momentum balance:
\begin{eqnarray*}
 / \!\!\! \vec{p}_{T} & = & E_{\nu_1} \cdot \vec{n_1} + E_{\nu_2} \cdot \vec{n_2}
\end{eqnarray*}
where $\vec{n_1}$ and $\vec{n_2}$ are directions of the two tau
candidates in the transverse plane (see right plot in
Fig.~\ref{event}).
While this is the simplest possible method to correct for the missing
neutrino energy, its clear advantage is that the solution is unique.

\begin{figure}
\includegraphics[width=0.5\textwidth]{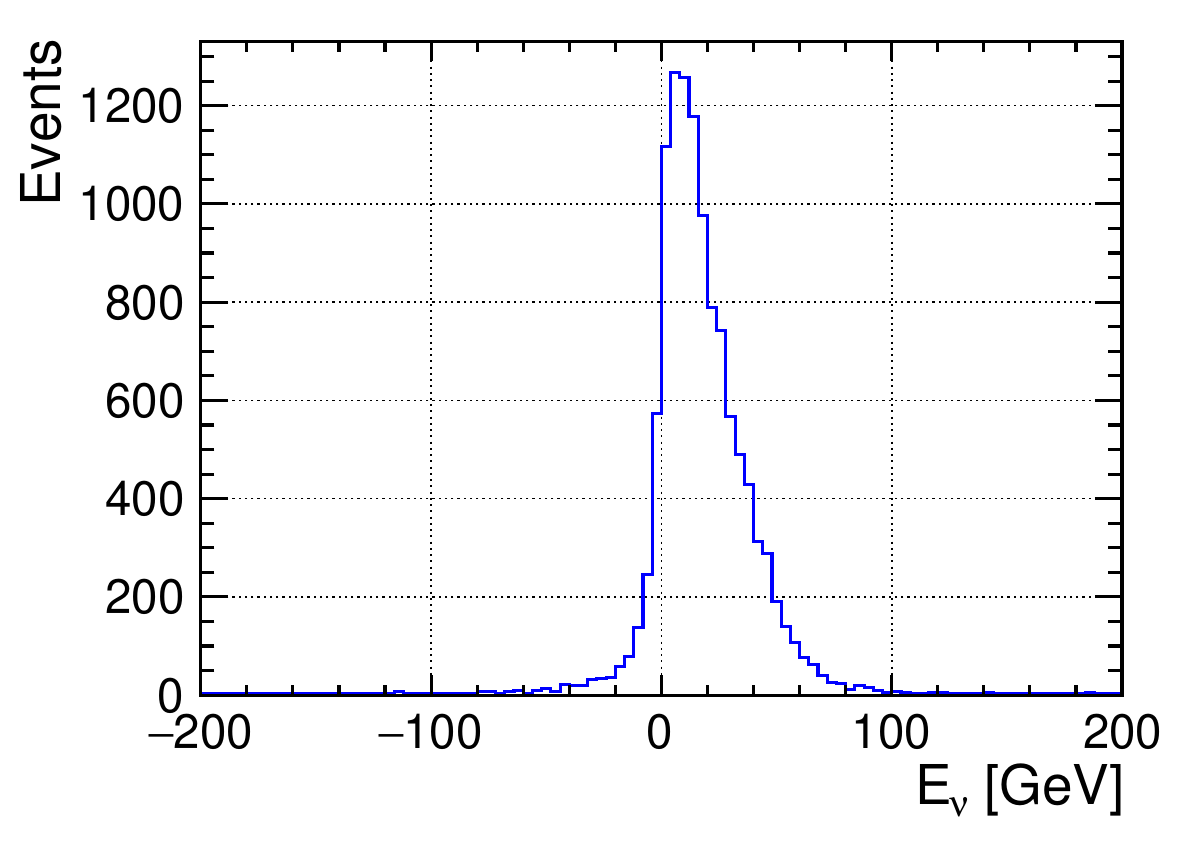}%
\includegraphics[width=0.5\textwidth]{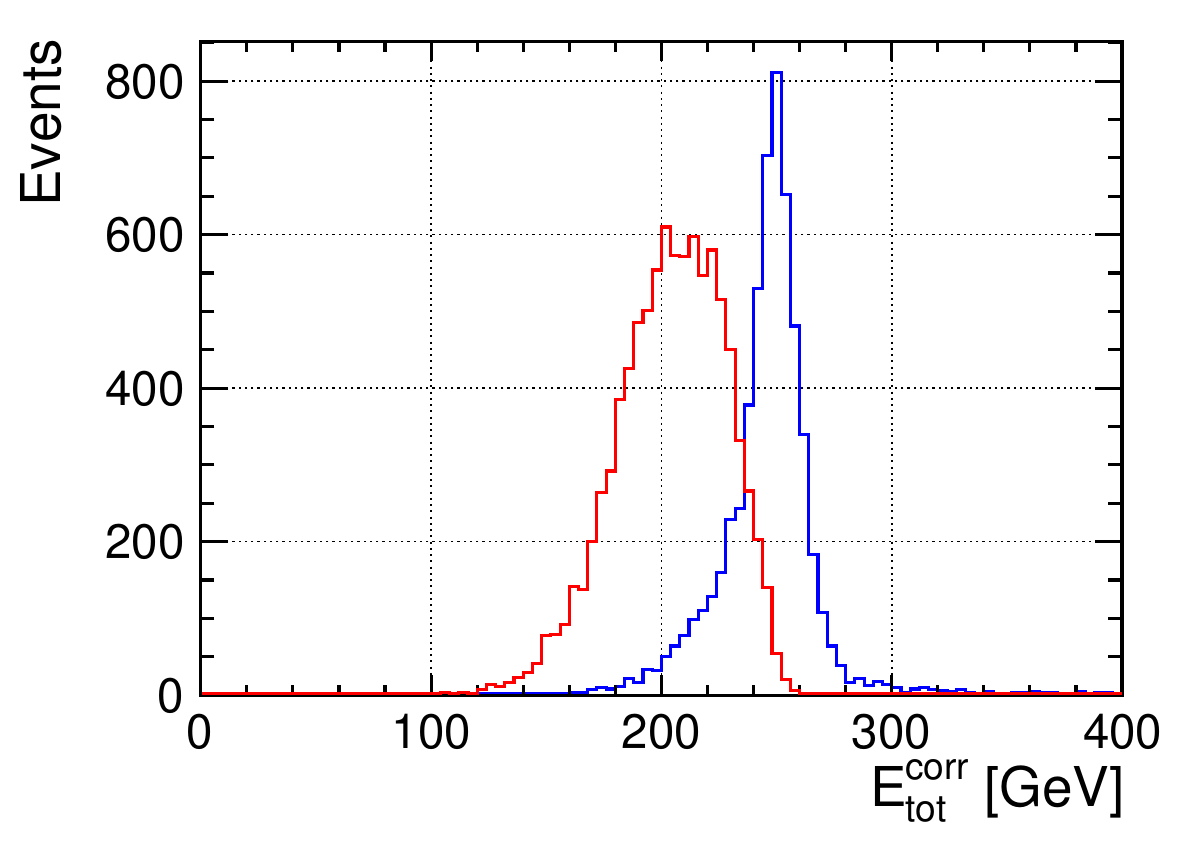}
\caption{Left: distribution of the neutrino energy for the
  signal events with hadronic tau decays, reconstructed from the
  transverse momentum balance.
  Right: Distribution of the total reconstructed energy of the event
  before (red) and after (blue) neutrino energy correction.
 }
\label{enu}
\end{figure}

Distribution of the reconstructed neutrino energy for the hadronic
signal events is shown in Fig.~\ref{enu} (left).
Due to finite detector resolution, negative neutrino energies are
reconstructed in a small fraction of cases. Neutrino emission is
neglected for given candidate in such cases.
When corrected for the reconstructed neutrino energies total reconstructed
energy of the event is in a good agreement with the expected collision
energy, see Fig.~\ref{enu} (right).

\begin{figure}
\includegraphics[width=0.5\textwidth]{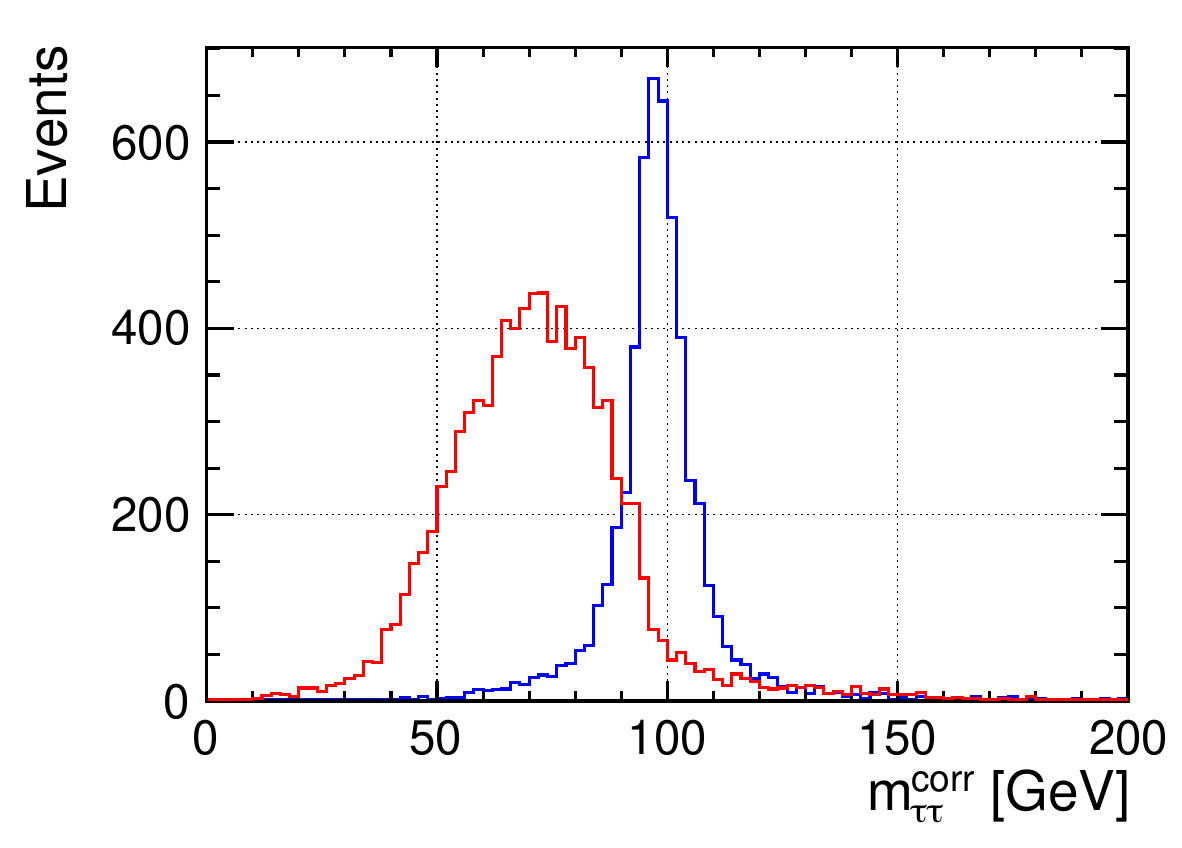}%
\includegraphics[width=0.5\textwidth]{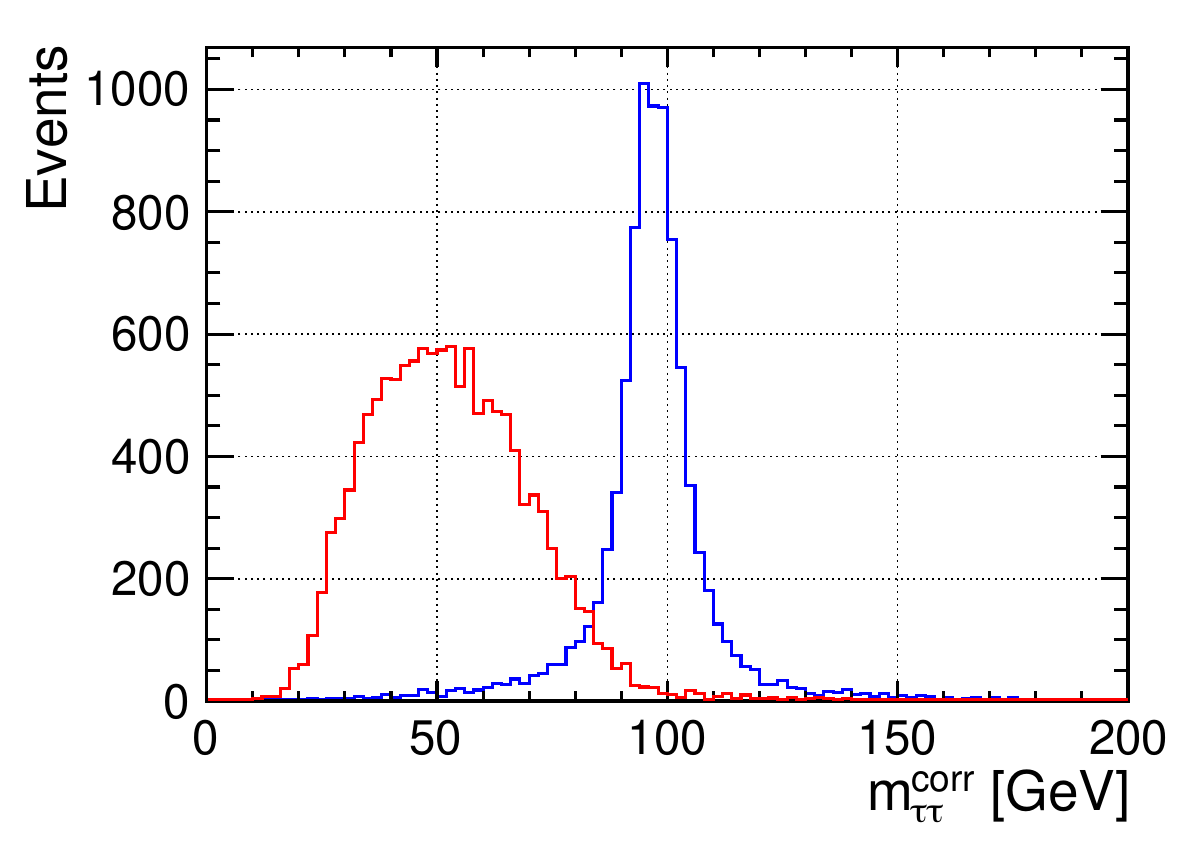}
\caption{Reconstructed invariant
  mass of the two tau candidates, before (red) and after (blue) the
  collinear energy correction. Distributions are shown for signal of
  light scalar production with mass of 95.7\,GeV,
  for hadronic (left) and semi-leptonic (right) event selection.  }
\label{mass}
\end{figure}

Compared in Fig.~\ref{mass} are the raw (before correction) and
corrected invariant mass distributions of the tau candidate pairs in
signal events.
It turns out that the collinear correction allows to reconstruct the
scalar mass with about 5\,GeV precision not only for hadronic events (left
plot) but also for semi-leptonic (right plot) and leptonic events.
This is because the invariant mass of the two neutrinos emitted in the
leptonic tau decay has to be small (below tau mass), so neglecting it
does not bias the estimate of the escaping energy significantly.

\section{First results}

\begin{figure}
\includegraphics[width=0.5\textwidth]{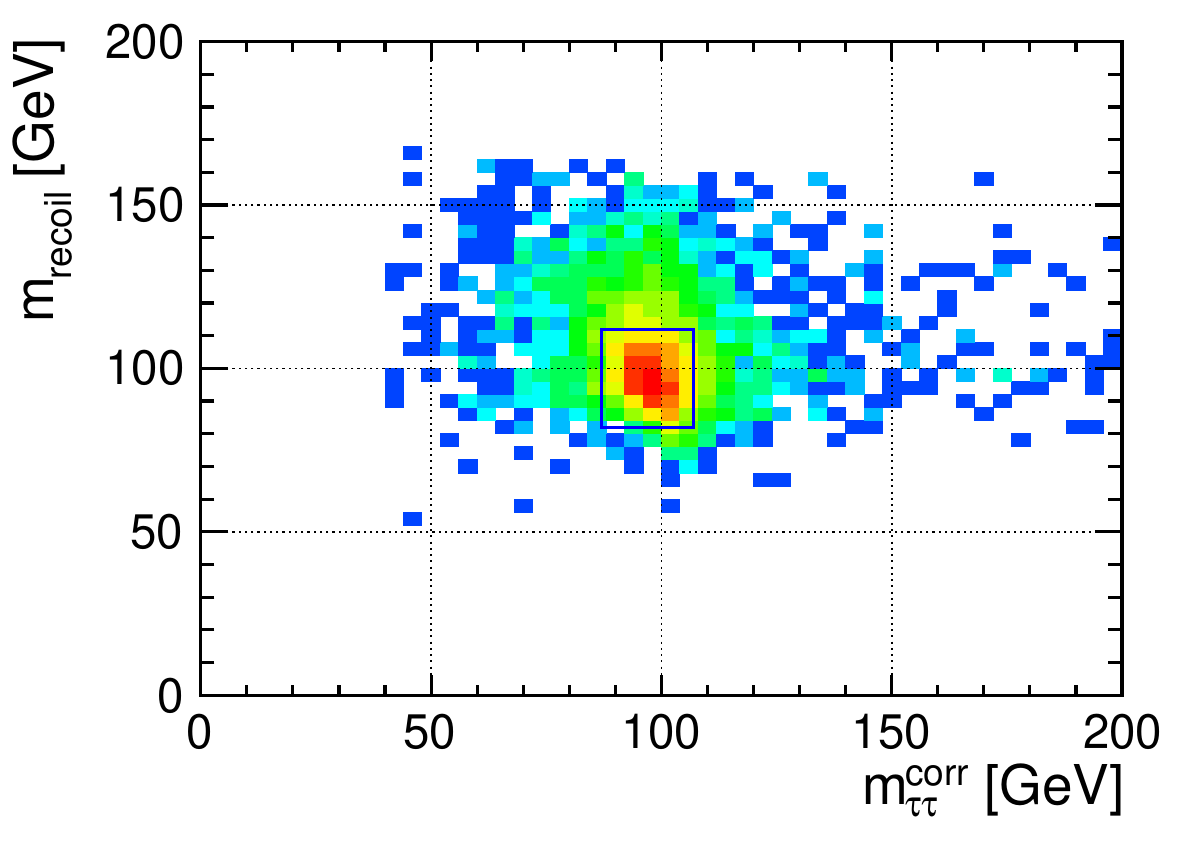}%
\includegraphics[width=0.5\textwidth]{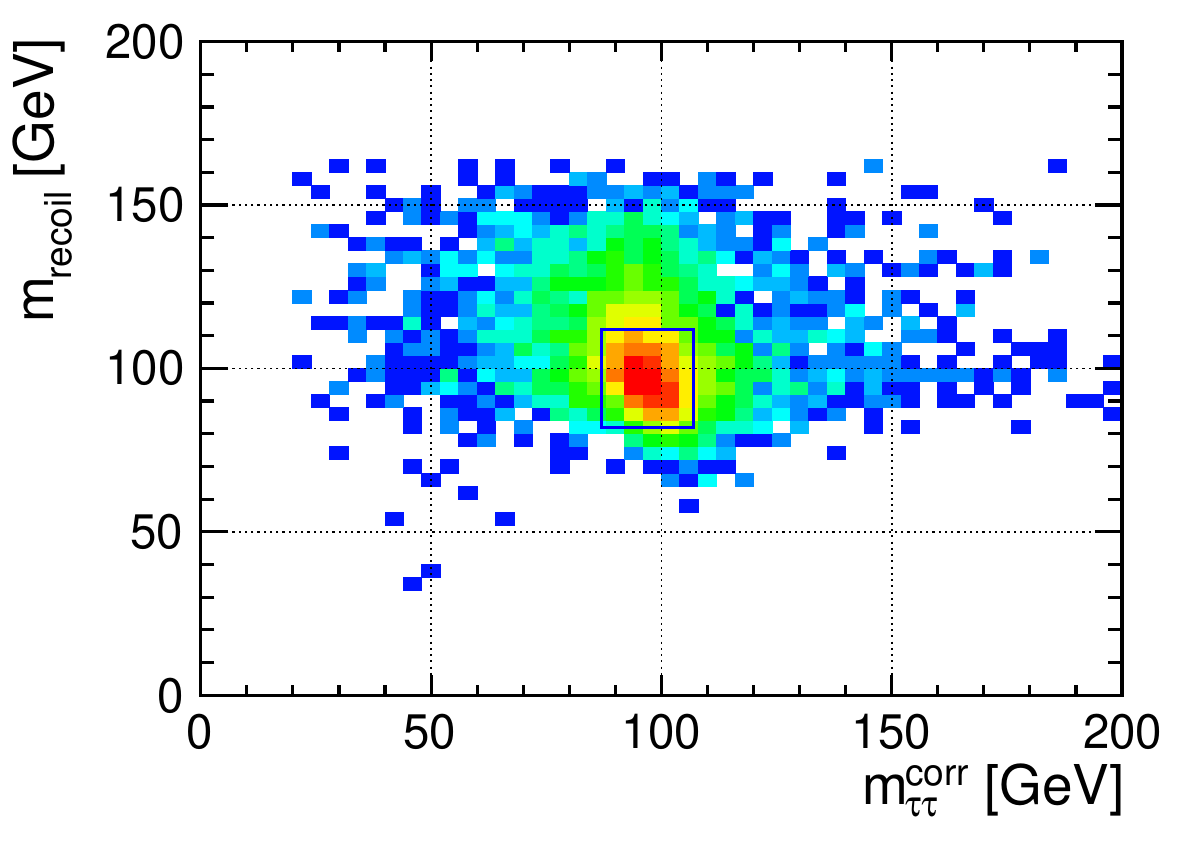}

\includegraphics[width=0.5\textwidth]{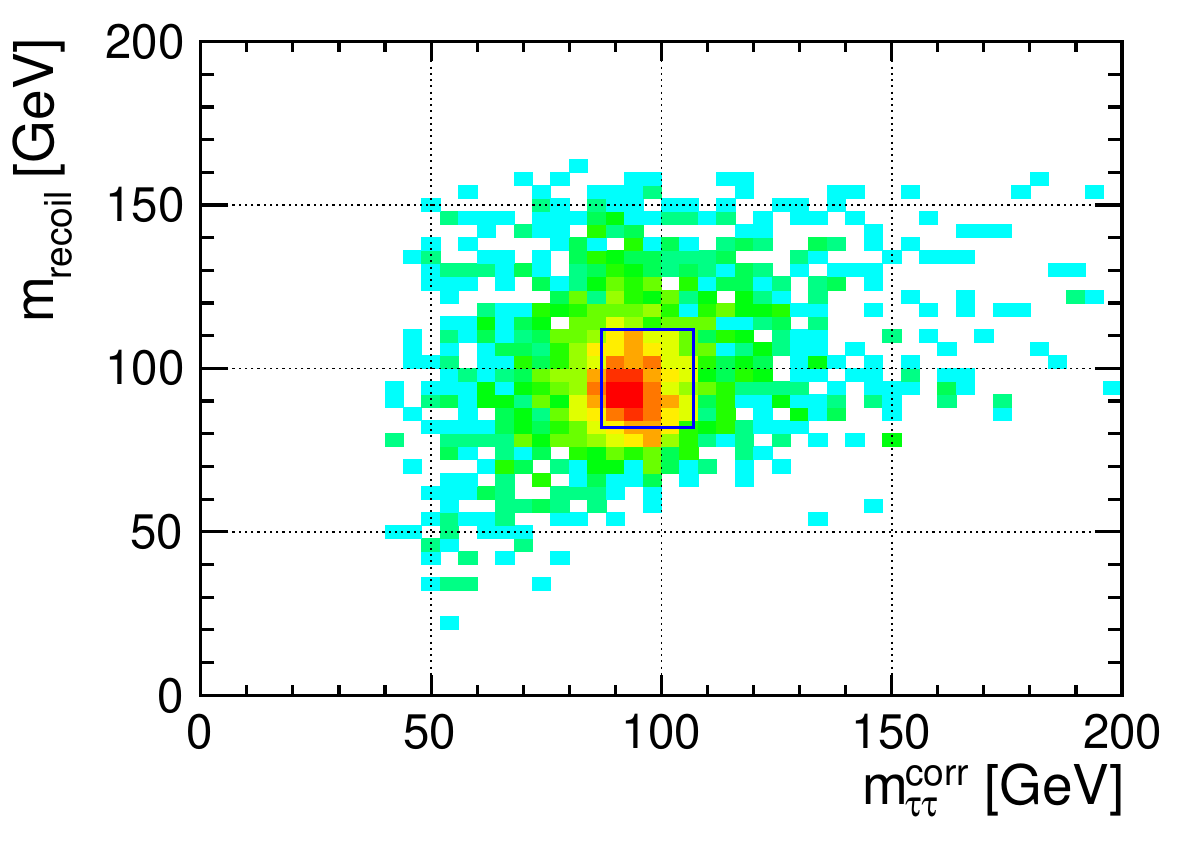}%
\includegraphics[width=0.5\textwidth]{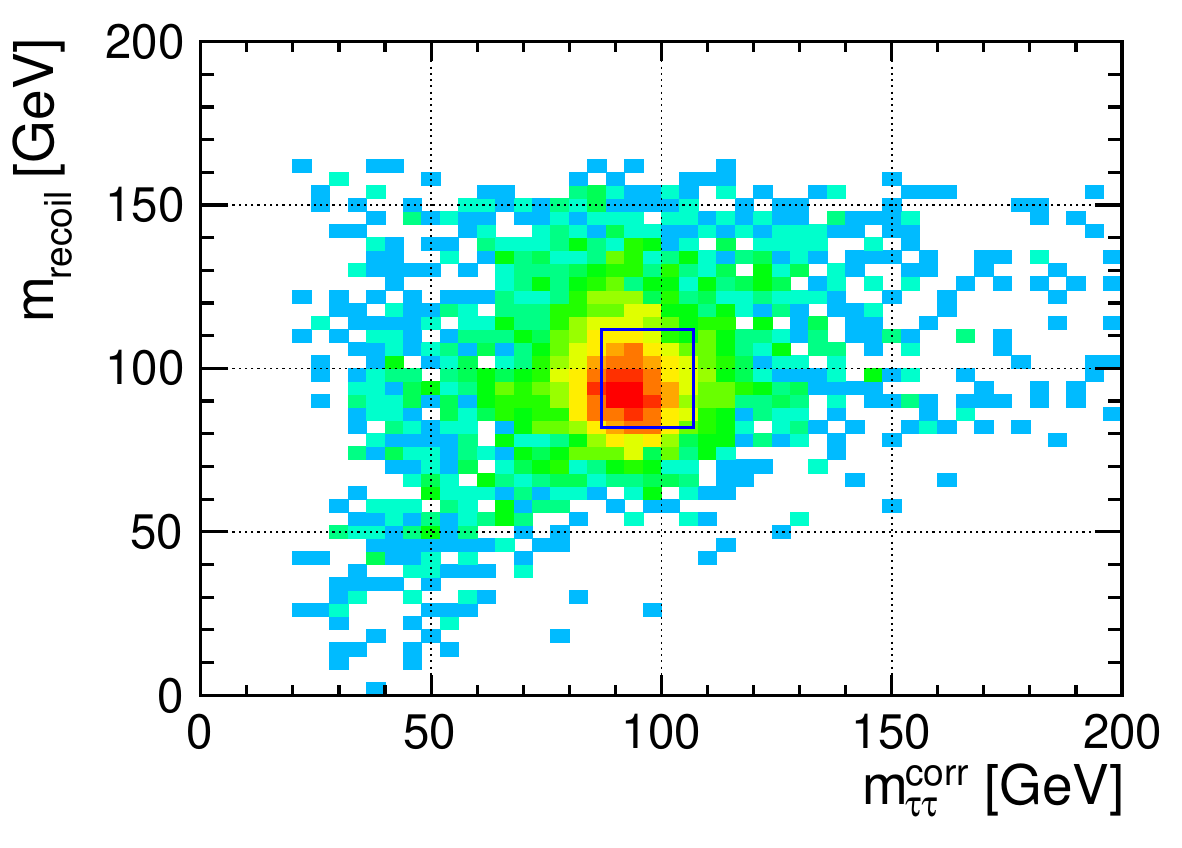}
\caption{Mass of the final state recoiling against the reconstructed
  Z boson decay, m$_\textrm{recoil}$, as a function of the corrected invariant
  mass of the two tau candidates, m$_{\tau\tau}$. Compared are expected
  distributions for signal of light scalar production (top)
  and dominant SM background from $e^+e^- \to q\bar{q} \tau \tau$
  (bottom), for hadronic (left) and semi-leptonic (right) event
  selection. See text for details.
  } 
\label{mass_2d}
\end{figure}

For the first estimate of the ILC sensitivity to the considered
benchmark scenario we used the simple cut-based approach.
After selecting events in different decay channels, four window cuts were
applied:
\begin{itemize}\setlength\itemsep{-0.5em} 
\item on the invariant mass of di-jet from Z decay,
\item on the raw invariant mass of two tau candidates,
\item on the corrected invariant mass of the two taus,
\item on the recoil mass (scalar mass reconstructed from the measured
  Z boson based on the energy-momentum conservation).  
\end{itemize}
Cuts applied on the di-jet invariant mass and the uncorrected
invariant mass of the tau candidate pair are indicated as blue boxes
in Fig.~\ref{rawmass_2d}. 
Distributions of the  corrected invariant mass of the two taus and the
recoil mass, with corresponding cuts indicated, are shown in
Fig.~\ref{mass_2d}.
After the neutrino energy correction, distributions for hadronic and
semi-leptonic events look very similar.
Also included in Fig.~\ref{mass_2d} (bottom plots) are the expected
background distributions, for the dominant background channel $e^+e^-
\to q\bar{q}\tau\tau$, mainly resulting from Z pair production.
Distribution for the background events is slightly shifted towards
lower masses of the tau pairs, but the difference between the
considered scalar mass of 95.7\,GeV and Z boson mass of 91.2\,GeV is
too small to allow for efficient background rejection.
On the other hand, other considered background channels are
efficiently rejected with the proposed set of cuts.

\begin{table}
  \begin{center}
    \def\arraystretch{1.3}
\begin{tabular}{|l|c|c|c|}
\hline\hline
Decay    & \multicolumn{2}{c|}{Events expected} & Signal       \\[-1mm] \cline{2-3}
channel  & Signal          &          Total bg. & significance  \\
\hline
Hadronic &  424 & 2730 & 7.5 \\ \hline
Semi-leptonic &  702 & 3490 & 10.8 \\ \hline
Leptonic &  260 & 1350 & 6.5 \\ \hline
\hline
\end{tabular}
\end{center}
  \caption{Summary of the results from the cut-based analysis.
  Significance of the signal of light scalar production with decay to
  two tau leptons is shown for the three considered decay channels.
  }
  \label{tab}
\end{table}

Summary of the results from the cut-based analysis is presented in
Tab.~\ref{tab}.
The signal to background ratio of about 0.15 was obtained for the
hadronic final state, while ratios of about 0.2 were estimated for
semi-leptonic and leptonic event samples. 
Stronger suppression of the dominant background channel,  $e^+e^- \to
q\bar{q} \tau \tau$, can probably be obtained with use of more
advanced analysis approaches based on machine learning.
Still, even in this very simple approach, using one polarisation
running only, signal of new scalar production in the considered
benchmark scenario should be visible at ILC with very high significance.

\section{Conclusions}

Beyond the SM scenarios with light scalars are still not excluded by
the existing data.
Sizable production cross sections for new scalars are still possible,
if combined with non-standard decay patterns.
Decays to tau pairs for scalars with mass close to m$_\textrm{Z}$ seem a challenging
benchmark scenario and a good testing ground for detector design and
analysis methods.
Initial study based on the N2HDM scenario, with a new scalar of
95.7\,GeV decaying to tau pairs, indicate that the considered signal
should be detectable at ILC running at 250\,GeV with high significance.
Experimental sensitivity should significantly increase when more
advanced analysis methods (MVA) are used.
The study will continue to get better understanding of signal
and background processes.

\subsection*{Acknowledgments}

We are very grateful to Thomas Biek\"otter for providing us with the
dedicated UFO file for the S2HDM modified for type IV couplings
and for his help in understanding the technical problems.

This work was partially supported by the National Science Centre
(Poland) under the OPUS research project no. 2021/43/B/ST2/01778.

\printbibliography

\end{document}